\documentclass[a4paper,11pt]{article}
\pdfoutput=1 % if your are submitting a pdflatex (i.e. if you have
\usepackage{jheppub} % for details on the use of the package, please
\usepackage[T1]{fontenc} % if needed

\usepackage{graphicx}% Include figure files
\usepackage{dcolumn}% Align table columns on decimal point
\usepackage{bm}% bold math

\usepackage{epstopdf}
\usepackage{epsfig}
\usepackage{graphics}
\usepackage{xspace}
\usepackage{slashed}
\usepackage{feynmp}
\usepackage{ams fonts}
\usepackage{sidecap}
\usepackage{longtable} % long tables
\usepackage{booktabs}
\usepackage{color}
\usepackage{subfigure}% in preamble
\usepackage[section]{placeins}
\usepackage{subfigure}
\usepackage{float}

\hypersetup{pdfpagemode=FullScreen}

\def\OMIT#1{}

\newcommand{\nn}{\nonumber}

\newcommand{\bea}{\begin{eqnarray}}
\newcommand{\eea}{\end{eqnarray}}

\usepackage{dcolumn} % Align table columns on decimal point
\usepackage{longtable} % long tables
\usepackage{booktabs}
\usepackage{tabularx}

\usepackage{float}

\title{\boldmath Unphysical Poles of Domain Wall Fermions at finite $L_s$} 

\collaborationImg{\includegraphics[width=.20\textwidth,origin=c]{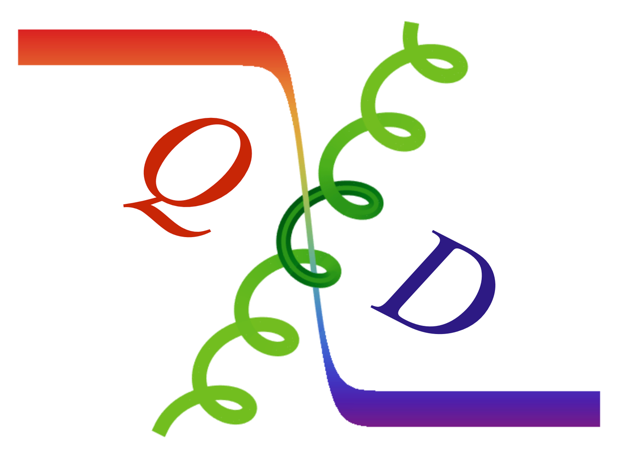}}
\collaboration{$\chi_{QCD}$ Collaboration}
\author[a]{Raza Sabbir Sufian,}
\author[a]{Michael J. Glatzmaier,}
\author[a]{and Yi-Bo Yang}

\affiliation[a]{Department of Physics and Astronomy, University of Kentucky, Lexington, KY 40506}
\emailAdd{sabbir.sufian@uky.edu}
\emailAdd{michael.glatzmaier@gmail.com}
\emailAdd{ybyang@pa.uky.edu}

\abstract{
We investigate the origin and behavior of oscillations observed in the hadron correlators constructed from the domain wall fermion (DWF) for different parameters involved in lattice QCD simulations. This oscillatory behavior at early time slices hinders the extraction of excited states in hadron spectroscopy.  Furthermore, the deviation from exponential decay may have a significant impact on fermion loop calculations performed on the lattice. We present results for several well-known implementations of the DWF actions. We extend the study of Shamir DWF action to include Bori\c{c}i and M\"{o}bius DWF actions by analyzing the poles of 4D quark propagator. For each action considered, we find an unphysical mode when analyzing the pole structure of the free DWF propagator for a finite extent of the $5^{\text{th}}$ dimension $L_s$, and we show that this mode is responsible for the oscillatory behavior observed in hadron correlators.  We have performed numerical checks on these results and have found that the presence of oscillatory behavior is sensitive to the DWF parameters $a_5$, $b_5$, $c_5$ and DW height $M$.  To minimize oscillations, our results suggest that one should choose $M a_5<1$ for the Shamir and Bori\c{c}i DWFs, and $M(b_5-c_5)<1$ for M\"{o}bius DWF when $M>1$.  For each calculation considered, the Bor\c{c}i DWF displayed the smallest magnitude of oscillation when compared to the Shamir and M\"{o}bius DWF actions using the same input parameters.
}

\begin{document} 
\maketitle
\flushbottom

\section{Introduction}
\label{sec:intro}

Formulating lattice fermions with exact chiral symmetry at finite lattice spacings is a nontrivial task.  One such realization of chiral symmetry is the domain wall fermion (DWF) on a (4+1)-dimensional lattice~\cite{Kaplan, Shamir}. Domain wall fermions have been used in several collaborations around the world (RBC/UKQCD, LHPC, TWQCD) to perform large-scale dynamical simulations of lattice QCD due to its good chiral symmetry. In many of these simulations, however, an unphysical oscillatory mode appears in hadron correlators within the first several time slices.  This unphysical mode has been observed when analyzing two point correlation functions computed on quenched configurations~\cite{Dudek}, in DBW2 action~\cite{Aoki}, LHPC hybrid action using DW fermions on HYP smeared MILC configurations~\cite{Hagler}, domain wall valence on Asqtad staggered sea configurations~\cite{Walker}, and domain wall valence on domain wall sea~\cite{Aoki2}.  A numerical analysis performed in~\cite{Negele} has shown that oscillation effects appear for domain wall heights $M>1$ and worsen as $M$ is increased for the Shamir action. 

A more recent analysis~\cite{Ying} of the Shamir action  has identified the unphysical mode responsible for the oscillatory behavior by analyzing the poles of 5D free DWF propagator~\cite{Shamir, xu}. They have shown, this unphysical mode acquires an imaginary part $i\pi$ when $M>1$ and results in oscillatory behavior of the DWF propagator in time.

In this work, we pinpoint the origin of these oscillatory modes for the Shamir, Bori\c{c}i, and M\"{o}bius DWF actions~\cite{Shamir,Borici,Brower} by analyzing the poles of 4D quark propagator.  We show that oscillatory effects appear for specific parameter choices in the free 4D propagator for each action.  We do this both analytically, by computing the tree-level pole terms for each DWF action, and then confirm our findings numerically using Chroma~\cite{chroma}.  We identify the pole responsible for the oscillatory behavior and refer to this pole as the \textit{unphysical pole} for the rest of this work.  For our numerical calculations, we plot the effective mass of the nucleon as a function of time for each action. The Bori\c{c}i DWF exhibits the smallest observable oscillation in the transfer matrix. We discuss this observation by noting that the
contribution of the 5D unphysical eigenmode of the Bori\c{c}i DWF transfer matrix has minimal coupling to the 4D boundary, and thus has minimal impact on the physical 4D propagator. We also suggest choices of the parameters $a_5$, $b_5$, $c_5$ for $M>1$ to minimize the oscillations in the hadron correlators.
% mjg - suggestion : the phrase ``provide a possible explanation'' sounds like we are scared to explain our analysis.  I suggest we replace that phrase with ``we discuss'' or just ``explain''.

This paper is organized as follows, an overview of the DWF actions considered as well as common notation is presented in sec.~(\ref{dwfs}). In secs.~(\ref{shamir_dwf}), (\ref{borici_dwf}), (\ref{mobius_dwf}) we present an analysis of the Shamir, Bori\c{c}i, M\"{o}bius DWFs and outline the method we use to extract the pole terms for each action. In sec.~(\ref{discuss_oscillation}), we show the effective mass plots of nucleons computed from Chroma for each DWF action using different sets of values of 5D lattice spacing $a_5$, $b_5$, $c_5$, domain wall height $M$ and extent of $5^{\text th}$ dimension $L_s$.  Finally, we examine the contributions of the unphysical eigenmodes at the 4D boundary for each DWF action and discuss why the oscillation in the Bori\c{c}i DWF is observed to be weaker than the Shamir and M\"{o}bius DWF case.  The detailed analytic formulas for the Bori\c{c}i 4D propagators can be found in Appendix~\ref{apxA} of the paper.

%
%%%%%%%%%%%%%%%%%%%%%%%%%%%%%%%
%  DWF
%%%%%%%%%%%%%%%%%%%%%%%%%%%%%%%
%

\section{The Domain Wall Fermion}\label{dwfs}
Domain wall fermions preserve flavor symmetry and they have greatly reduced chiral symmetry breaking at the expense of adding an extra fifth dimension. The four dimensional chiral fermions arise on the boundaries of a five dimesional lattice. 
In this section we outline several of the actions used to compute the free tree-level propagator using DWFs.  We make use of the compact notation in~\cite{Brower} by keeping the length of the $5^{\text{th}}$ domain wall dimension $L_s=4$,
\bea
D^{\text{DWF}}_{s,s'}=
\left(\begin{array}{cccc}
D_+^1 & D_-^1\>P_L & 0 & -mD_{-}^{1}P_R       \\
D_-^2P_R & D_+^2 & D_-^2P_L & 0      \\
0 & D_-^3P_R & D_+^3 & D_-^3P_L       \\
-mD_{-}^{4}P_L & 0 & D_-^4P_R & D_+^4
\end{array}\right) \nn \\
\eea

Where in these equations we have used $s$ and $s'$ as row and column indices respectively, $m$ is the bare quark mass and we have adopted the shorthand,
\bea \label{gen}
D_+^{s} &=& b_s D_W +1; \qquad D_-^{s}=c_sD_W -1;\qquad  P_{R,L} = \frac{1\pm \gamma_5}{2} 
\eea
Terms proportional to the quark mass $m$ are boundary condition terms ($s=1$ or $s=L_s$).  Each action we consider corresponds to differing values of the $b_s$ and $c_s$ coefficients, these are 
\cite{Brower},
%\begin{table}[tbp]
\begin{table}[H]
\centering   
%\begin{ruledtabular}
\begin{tabular}{|lr|c|}
\hline
 Action & $b_s$ & $c_s$ \\
 \hline
 Shamir & $a_5$ &  $0 $ \\
 Bori\c{c}i & $a_5 $ &  $a_5  $ \\
 M\"{o}bius & $b_5$ &  $c_5$ \\
 \hline
\end{tabular} 
%\end{ruledtabular}
  \caption{The values of $b_s$ and $c_s$ for the various DWF actions considered.}
  \label{bscs}
\end{table}

\noindent For tree-level calculations in the momentum space, we set gauge links to unity, $U_{\mu}\to 1$ and the Wilson parameter $r_W=1$.  We write the Dirac-Wilson operator in the momentum space with a negative mass term $M$ as,
\bea \label{wils}
D_W(p) = -M+\frac{i}{a}\sum_{\mu}\gamma_{\mu}\sin p_{\mu}a + \frac{1}{a}\sum_{\mu}(1-\cos p_{\mu}a)
\eea
Using this form for $D_W(p)$, we write the Dirac operator multiplied by $a_5$ as,
{\small
\bea \label{M}
a_5D_W(p)&=&\overline{a}_5\left(-\overline{M}+{i}\sum_{\mu}\gamma_{\mu}\sin p_{\mu}a + \sum_{\mu}(1-\cos p_{\mu}a)\right)\nn \\
&\text{where}& \qquad \overline{a}_5\equiv\frac{a_5}{a} \qquad\text{and}\qquad  \overline{M}\equiv Ma
\eea}
\newline

%
%%%%%%%%%%%%%%%%%%%%%%%%%%%%%%%%%%
%
%            SHAMIR POLES 
%
%%%%%%%%%%%%%%%%%%%%%%%%%%%%%%%%%%

\section{Poles of the Shamir Domain Wall Fermion}\label{shamir_dwf}
In this section we re-derive several known results for the Shamir DWF action.  In particular, we confirm that when the domain wall height $M>1$, the 5D transfer matrix becomes complex~\cite{shamir2} which gives rise to oscillatory behavior in the hadron correlators.
%From various numerical results, like ~\cite{blum}, for optimal value of $M$ between $1.6$ and $1.7$ for $\beta=6$,  the transfer matrix of Shamir DWFs is complex.
We begin by considering the Shamir DWF operator and take $b_s=a_5$ and $c_s=0$ in eq.~(\ref{gen}). To construct 5D Green's function we follow the method described in~\cite{Shamir} and first consider the Dirac operator on an infinite $s$-direction,
\bea
D^0_{s,s'} = (a_5 D_w +1)\delta_{s,s'} - P_L\delta_{s,s'-1} - P_R\delta_{s,s'+1}.
\eea
We then compute the second order operator $\Omega_{ss'}^0 = \sum_t D^0_{s,t}D_{t,s'}^{0\dagger}$ and then compute $G_{s,s'}^0 =\left(\Omega_{s,s'}^0\right)^{-1}$.  With these comments, we begin by using the form of $D_W$ given in eq.~(\ref{wils}) and find for $\Omega^0_{ss'}$,
\bea
\Omega^{0}_{s,s^{\prime}}&&=\sum_t \ D^0_{s,t}D_{t,s'}^{0\dagger} \nonumber \\
&&=2b(p) ( \cosh \alpha(p)\delta_{s,s^{\prime}}-\frac{1}{2} (\delta_{s,s^{\prime}+1}+\delta_{s,s^{\prime}-1}) )
\eea
where,
\bea\label{cosh}
&&b(p)=\overline{a}_5(-\overline{M}+\sum_{\mu}(1-\cos p_{\mu}))+1,\nn\\
&&\overline{p}^2 = \sum_\mu\sin^2 p_\mu a,\>\>\>\ \cosh\alpha(p)=\frac{\overline{a}_5^{2}\overline{p}^{2}+b^{2}(p)+1}{2b(p)}
\eea
The inverse of $\Omega^{0}$ is given by, Refs.~\cite{Shamir,Aoki3},
\bea
G^{0}_{s,s^{\prime}}=\left(\Omega^{0}_{s,s^{\prime}}\right)^{-1}= A_0e^{-\alpha|s-s'|}
\eea
where,
\bea
A_0=\frac{1}{2b(p)\sinh\alpha}
\eea
One can verify by explicit calculation that, for $s=s^{\prime}$,
\bea
[\Omega^{0}]_{s,t}[G^{0}]_{t,s^{\prime}}=\mathbb{I}_{s,s'}
\eea
and is zero otherwise.

Following~\cite{Shamir}, we now consider finite $5^{\text{th}}$ dimension $0\leq s\leq L_s$ and turn on link connecting sites $s=0$ and $s=L_s$, therefore two Weyl fermions will now form Dirac fermions and their mass is now proportional to the strength of the link. With these results we can now compute the 5D Green's function for finite $L_s$. In what follows, we also scale $a_5=a=1$ in the intermediate steps of the calculation and restore them at the end as was done in~\cite{Shamir, Aoki5, xu} to simplify the manipulations. We follow the conventions in~\cite{Aoki5}, i.e. we take $L_s \to L_s-1, s \to s-1, s' \to s'-1 $ and calculate the full 5D Green's function following the method in~\cite{Shamir, Aoki5},
\bea
G^{L,R}_{s,s'}=&& A_0e^{-\alpha|s-s'|}+A_{\mp}e^{-\alpha(s+s'-2)}+A_{\pm}e^{-\alpha(2L_s-s-s')} \nonumber \\
&&+A_m\left(e^{-\alpha(L_s-s+s'-1)}+e^{-\alpha(L_s+s-s'-1)}\right)
\eea
where we have defined,
\bea
A_0=&& \frac{1}{2b(p)\sinh\alpha} \nn \\
A_-=&&-\frac{A_0}{F_N}(1-m^2)(1-b(p)e^{-\alpha}) \nn \\
A_+=&&-\frac{A_0}{F_N}(1-m^2)(1-b(p)e^{\alpha})e^{-2\alpha} \nn \\
A_m=&&-\frac{A_0}{F_N}e^{-\alpha}(2mb(p)\sinh\alpha -e^{-\alpha L_s}((1-b(p)e^{-\alpha}) -m^2(1-b(p)e^{\alpha}))) \nn \\
F_N=&&(1-b(p)e^{\alpha})-m^2(1-b(p)e^{-\alpha})+e^{-\alpha L_s}(4\sinh\alpha (1-m^2)) \nn \\ &&
 -e^{-2\alpha L_s}((1-b(p)e^{-\alpha}) -m^2(1-b(p)e^{\alpha})) \nn
\eea
And we will make use of the usual definition of the residual mass,
\bea 
m_r=e^{-\alpha L_s}
\eea
With the above results for the 5D Green's function, we now project this 5D Green's function to 4D at finite $L_s=N$ following the method described in~\cite{Aoki4}. By using the fact that for any DWF action, the physical quark fields in 4D are defined on the boundaries of the $5^{\text{th}}$ dimension ($s=1$ and $s=N$),

\bea \label{q1}
q(x)=P_R\psi_1(x)+P_L\psi_N(x)\nonumber \\
\bar{q}(x)= \bar{\psi}_1(x)P_L+\bar{\psi}(x)_NP_R
\eea
Starting with the 5D propagator, we can use the above relations to compute the corresponding 4D propagator,
\bea \label{project-4d}
S^{F,5D}_{s,s'}(p) &=& \langle\psi_s(-p)\>\bar{\psi}_{s'}(p)\rangle\\
S^{4D}_F(p) &=& \langle q(-p)\>\bar{q}(p)\rangle\nn\\
&=&P_L\langle\psi_1(-p)\>\bar{q}(p)\rangle +P_R\langle \psi_N(-p)\>\bar{q}(p)\rangle
\eea
For finite $L_s$ the physical 4D propagator takes the form,
\bea \label{phyprop}
S^{4D}(p)=
&& -i\slashed{\bar{p}}\Bigg( \frac{m_r^2[(1-b(p)e^{-\alpha})-m^2(1-b(p)e^{\alpha})]}{BF_N}-
\frac{m_r^2(1-m^2)(1-b(p)e^{\alpha})}{BF_N}\nn \\ &&-\frac{(1-m^2)(1-b(p)e^{-\alpha})}{BF_N}-
 \frac{2mm_r}{F_N}+\frac{1}{B} 
+\frac{m_r^2[(1-b(p)e^{-\alpha})-m^2(1-b(p)e^{\alpha})]}{BF_N} \Bigg) \nn \\
&&-b(p)\Bigg(\frac{m_re^{\alpha}}{B}-\frac{m_re^{-\alpha}(1-m^2)(1-b(p)e^{\alpha})}{BF_N}-\frac{mm_r^2e^{\alpha}}{F_N}\nn\\
&&+\frac{m_r^3e^{\alpha}[(1-b(p)e^{-\alpha})-m^2(1-b(p)e^{-\alpha})]}{BF_N} \Bigg)+m\Bigg( \frac{m_r^2[(1-b(p)e^{-\alpha})-m^2(1-b(p)e^{\alpha})]}{BF_N} \nn \\
&& -\frac{m_r^2(1-m^2)(1-b(p)e^{\alpha})}{BF_N}+\frac{1}{B}-\frac{(1-m^2)(1-b(p)e^{-\alpha})}{BF_N}-\frac{2mm_r}{F_N}\nn \\
&&+\frac{m_r^2[(1-b(p)e^{-\alpha})-m^2(1-b(p)e^{\alpha})]}{BF_N} \Bigg)
\eea
where we have defined, 
\bea
B=A_0^{-1}=2b(p)\sinh\alpha
\eea
We shall show in the following sections that the $B$-term contributes to the unphysical pole (e.g. it gives rise to oscillatory contributions for $M>1$) and $F_N$ contributes to the physical pole of the propagator. As $L_s\to \infty$, we can neglect all the terms proportional to $e^{-\alpha L_s}$ and one can show that the unphysical pole disappears from the theory and reduces to the propagator in~\cite{Aoki4}. Then,
\bea
&&S^{4D}(p)\big\vert_{L_s\to\infty}=\nn \\
&&-i\slashed{\bar{p}}\frac{[(1-b(p)e^{\alpha})-m^2(1-b(p)e^{-\alpha})]-(1-m^2)(1-b(p)e^{-\alpha})}{B[(1-b(p)e^{\alpha})-m^2(1-b(p)e^{-\alpha})]}+\nn \\
&&\frac{Bb(p)me^{\alpha}+m[(1-b(p)e^{\alpha})-m^2(1-b(p)e^{\alpha})]-m(1-m^2)(1-b(p)e^{-\alpha})}{BF} \nn \\
&&=i\slashed{\bar{p}}\frac{2b(p)\sinh\alpha}{BF}+\frac{b(p)me^{-\alpha}-m}{F} \nn \\
&&=\frac{i\slashed{\bar{p}}-m(1-b(p)e^{-\alpha})}{F}
\eea
where, 
\bea
F=F_N\big\vert_{L_s\to \infty}=(1-b(p)e^{\alpha})-m^2(1-b(p)e^{-\alpha})
\eea
A short summary of these findings is presented in Table~\ref{poles_shamir}.

\begin{table}[H]
\centering
\begin{tabular}{|lr|c|}
\hline
$L_s$ limit & Tree-Level Pole Terms \\
\hline
$L_s$ finite  (4D) & $ \begin{cases}F_N \\
B\end{cases} $ \\ 
$L_s\to\infty$ (4D) &  $\>\>\>\>F=(1-b(p)e^{\alpha})-m^2(1-b(p)e^{-\alpha})$ \\
\hline
  %\label{spoletab}
\end{tabular} 
\caption{The various pole terms for the Shamir DWF action in both the finite and infinite $L_s$ limits.}
\label{poles_shamir}
\end{table}

%%%%%%%%%%%%%%%%%%%%%%%%%%%%%%%%%%
%
%            SHAMIR OSCILLATION RESULTS
%
%%%%%%%%%%%%%%%%%%%%%%%%%%%%%%%%%%

\subsection{Oscillation Effects in the Free Shamir Domain Wall Propagator}

It has been shown in ~\cite{Ying} that for $b(p)=0$ the second order operator $\Omega^0_{s,s'}$ is a unit matrix (up to a factor) whose inverse is $G^0_{s,s'}\propto \delta_{s,s'}$ and the corresponding modes are completely bound to the domain wall and are not responsible for the oscillation in the temporal direction of the hadron correlators. From our results in the previous section, the unphysical pole of the propagator in eq.~(\ref{phyprop}) can then be computed by solving for $p$ in the expression,
\bea
\sinh\alpha=0\nn
\eea
or equivalently from eq.~(\ref{cosh}),
\bea
\cosh\alpha(p)=\frac{\overline{a}_5^{2}\overline{p}^{2}+b^{2}(p)+1}{2b(p)} =1
\eea
To solve for $p$ in the above equation, we shall only consider a static mode, $p_{i} =0$, $p_4\neq 0$~\cite{Ying}.  We then find for $p_4$,
\bea \label{zeromode}
&&\overline{M}=1-\cos p_4a \mp i\sin p_4a\nonumber \\
&&\Rightarrow e^{\pm ip_{4}a}=1-Ma
\eea
where $\overline{M}=Ma$ defined in eq.~(\ref{M}). We then find that the energy, $E =- i p_4$,
\bea\label{shamir-condition}
E=\pm\frac{1}{a} \ln(1-Ma).
\eea
For $a=1$, $E$ is real for $0<M<1$. This implies that this static mode propagates forward or backward in the time direction with the energy $E=\ln(1-M)$.
However, in a simulation, if $ M > 1$
\bea\label{shamir:pole}
E=\pm\left[\ln(M-1) + i\pi\right]
\eea
which is complex. This unphysical mode propagates in the time direction as 
\newline $(-1)^t \left(e^{\ln(M-1)t}+e^{\ln(M-1)(T-t)}\right)$, where $\ln(M-1)$ is negative for $M>1$. This gives rise to oscillatory behavior in hadron correlation functions for Shamir DWF action.  Though it has been argued in~\cite{Negele} that these oscillations are due to the lattice artifacts at cutoff scales and in~\cite{Dudek, huey} that the non-locality of the valence DWF action in four dimensions produces oscillatory contributions to the two-point effective mass close to the source, this analysis illustrates that the unphysical pole term in eq.~\ref{phyprop} gives rise to oscillatory behavior in the DWF correlators.
%%%%%%%%%%%%%%%%%%%%%%%%%%%%%%%
%  Bori\c{C}I DWF Action
%%%%%%%%%%%%%%%%%%%%%%%%%%%%%%%
%
\section{Bori\c{c}i DWF Action}\label{borici_dwf}
In this section we consider Bori\c{c}i's action and set $a=1$ as was done in~\cite{Brower}. The DWF operator for Bori\c{c}i's action is given by setting,
\bea
b_s = a_{5}, \qquad c_s = a_{5}\nn
\eea
for the Dirac operators appearing in eq.~(\ref{gen}). With these values of $b_s$ and $c_s$ and following the calculation in sec.~(\ref{shamir_dwf}) we start with Dirac operator on an infinite $s$-direction, 
\bea
D^0_{s,s'}=&&\left(ia_5\overline{\slashed{p}}+b(p)\right)\delta_{s,s'}+\left(ia_5\overline{\slashed{p}}+c(p) \right)P_L\delta_{s,s'-1}\nn \\
&&+\left(ia_5\overline{\slashed{p}}+c(p) \right)P_R\delta_{s,s'+1}
\eea
where we have defined,
\bea
b(p)&=& a_5(-M + \sum_{\mu}(1-\cos p_{\mu}a))+1\nn \\
c(p)&=& a_5(-M + \sum_{\mu}(1-\cos p_{\mu}a))-1
\eea
Following a similar procedure as we did for the Shamir DWF, we compute $\Omega^0_{s,s'}=\sum_t D^0_{s,t}D_{t,s'}^{0\dagger}$.  For the Bori\c{c}i DWF we find for $\Omega_{s,s'}^0$,
\bea
\Omega^{0}_{s,s'}=&&2(-a_5^{2}\overline{p}^{2}-b(p)c(p))(\cosh\alpha^{\prime}(p)\delta_{s,s'}\nn \\
&&-\frac{1}{2}(\delta_{s,s'+1}+\delta_{s,s'-1}))
\eea
where,
\bea
\cosh\alpha^{\prime}(p)=\frac{2a_5^{2}\bar{p}^2+b^2 (p)+c^2 (p)}{2(-a_5^{2}\overline{p}^2-b(p)c(p))}
\eea
Using a similar ansatz as before for $\left(\Omega_{s,s'}^0\right)^{-1}$, we have,
\bea \label{borgreen}
G^{0}_{s,s'}=&&\frac{e^{-\alpha^{\prime}|s-s'|}}{2(-a_5^{2}\overline{p}^{2}-b(p)c(p))\sinh\alpha^{\prime}}\nn \\
=&&\frac{1}{B'}e^{-\alpha^{\prime}|s-s'|}
\eea
where
\bea
B'=&&2(-a_5^{2}\overline{p}^2-b(p)c(p))\sinh\alpha'
\eea
As before, one can verify through explicit calculation that this form of the inverse does indeed satisfy, for $s=s^{\prime}$,
\bea
[\Omega^{0}]_{s,t}[G^{0}]_{t,s^{\prime}}=\mathbb{I}_{s,s'}
\eea
and zero otherwise.

Following the method described in sec.~(\ref{shamir_dwf}) and having computed the 5D Green's function for finite $L_s$ using $G^0_{s,s'}$, we calculate the 4D quark propagator by taking $a,a_5=1$,~for simplification as was done for the Shamir DWF case. The full expression for the 4D propagator can be found in the Appendix~\ref{apxA} of the paper. Schematically, the 4D propagator can be written as,
\bea \label{borprop}
S^{4D}(p)=\frac{-i\slashed{\overline{p}}N-b'(p)N'+m'(p)N^{''}}{B'F'_N}
\eea

With these results then, we identify the $B'$ in the denominators in eq.~(\ref{borprop}) or (\ref{borgreen}) to give rise the unphysical pole in analogy with the Shamir DWF.  As was done with the Shamir DWF case, we extract the pole of the propagator setting $p_i=0$, and $p_4 \neq 0$ by solving the condition,
\bea
\sinh\>\alpha' =0\qquad\text{or}, \>\>\cosh\>\alpha'=1.
\eea

which results in a condition on $E=-ip_4 $ which is the same result as the Shamir DWF,
\bea
E=\pm \ln(1-M)
\eea
As before it is complex for $M>1$ and leads to the oscillatory behavior.
%%%%%%%%%%%%%%%%%%%%%%%%%%%%%%%
%  M\"{o}bius DWF Action
%%%%%%%%%%%%%%%%%%%%%%%%%%%%%%%
\section{M\"{o}bius DWF Action}\label{mobius_dwf}
In this section we consider the M\"{o}bius DWF operator.  The form for the DWF operator on an infinite $s$-direction is given by taking $a=1$, $b_s=b_5$, $c_s=c_5$, in eq.~(\ref{gen}), 
\bea
D^0_{s,s'}=&& (b_5(i\slashed{\overline{p}}+b(p))+1)\delta_{s,s'}+(c_5(i\slashed{\overline{p}}+b(p))-1)\nn \\ &&P_L\delta_{s,s'-1}+(c_5(i\slashed{\overline{p}}+b(p))-1)P_R\delta_{s,s'+1}
\eea 
where,
\bea
b(p)=-M+\sum_\mu (1-\cos p_\mu)
\eea
We find for $\Omega^0$ for the M\"{o}bius DWF,
\bea
\Omega^{0}_{s,s'}=&&2(b_5c_5(\overline{p}^2+b(p)^2)-b(p)(b_5-c_5)-1) \nn \\&&(\cosh\alpha^{''} (p)\delta_{s,s'}-\frac{1}{2}(\delta_{s,s'+1}+\delta_{s,s'-1}))
\eea
where
\bea
\cosh\alpha^{''}= \frac{(\overline{p}^2+b^2(p))(b_5^2+c_5^2)+2b(p)(b_5-c_5)+2}{2(1+b(p)(b_5-c_5)-b_5c_5(\overline{p}^2+b(p)^2))}\nn \\
\eea
Therefore, 
\bea
G^{0}_{s,s'}=\frac{e^{-\alpha^{''}|s-s'|}}{2(1+b(p)(b_5-c_5)b_5c_5(\overline{p}^2+b(p)^2))\sinh\alpha^{''}}. \nn \\
\eea
A calculation similar to the Shamir and Bori\c{c}i DWFs can be done to calculate the 4D quark propagator and one can show the unphysical pole is present analogous to Shamir and Bori\c{c}i 4D quark propagators. 
Solving for $\sinh\alpha^{''}=0$ or equivalently $\cosh\alpha^{''}=1$, we get 
\bea
(\overline{p}_4^2+b(p)^2)(b_5+c_5)^2=0 \nn \\
\Rightarrow M=1-e^{\pm i p_4}.
\eea
Again,
\bea
E=\pm \ln(1-M)
\eea
which is again complex for $M>1$ analogous to what we obtained for Shamir and Bori\c{c}i DWFs.

%
%%%%%%%%%%%%%%%%%%%%%%%%%%%%%%%
%  NUMERICAL RESULTS OF OSCILLATORY BEHAVIOR
%%%%%%%%%%%%%%%%%%%%%%%%%%%%%%%
%
\section{Numerical Results of Oscillatory Behavior}\label{discuss_oscillation}
\begin{figure}[h]
\begin{center} 
\includegraphics[height=8.0cm,width=10.0cm]{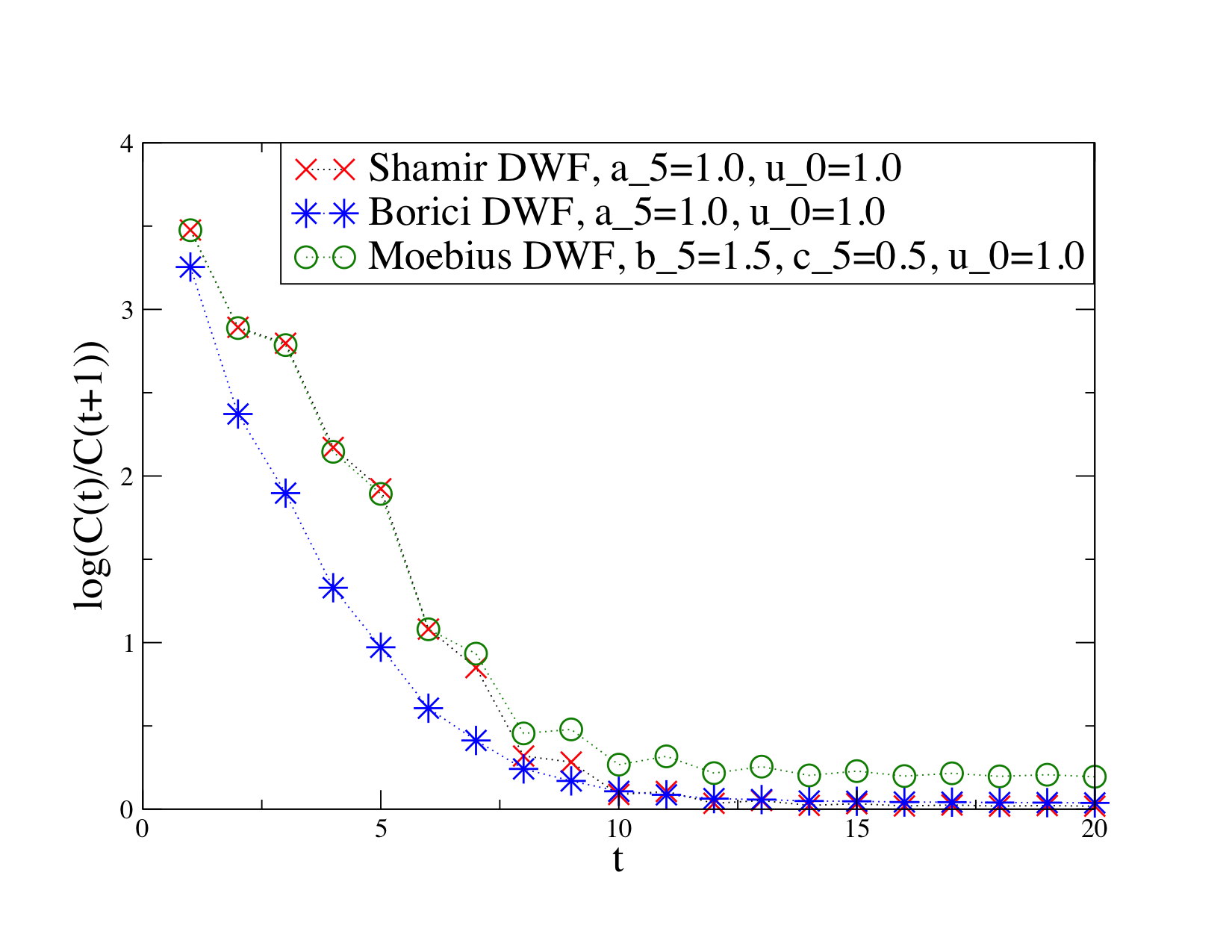}  
\caption{\small \sl $m_{eff}$ plot for $M=1.8$, $m=0.005$, $a_5=1.0$, $b_5-c_5=1.0$, $u_0=1.0$, $L_s=16.$\label{fig:1}}  
\end{center}  
\end{figure}

 In this section we compute hadron effective mass $m_{\text{eff}}$, and plot $m_{\text{eff}}$ for the proton as a function of time for various DWFs. We use Chroma~\cite{chroma} to construct various DWF correlators on free $24^3\times 64$ configurations for different values of $a_5$, $b_5$, $c_5$, $M$ and $L_s$.  We also consider configurations where the links have been replaced by their vacuum expectation value $u_0$~\cite{Lepage}.  We have used the standard definition of the effective mass in our plots,
\bea\label{meff}
m_{\text{eff}} &=& \log\>\frac{C(t)}{C(t+1)},
\eea
where $C(t)$ is the hadron correlator. Figure~\ref{fig:1} illustrates the oscillatory behavior at short time slices in the effective mass plots when $M=1.8$, $m=0.005$, $a_5=1$, $u_0=1$, $L_s=16$ for the free Shamir, Bori\c{c}i DWFs and $b_5=1.5$, $c_5=0.5$ for the M\"{o}bius DWFs in agreement with our analytic conditions.  

\begin{figure}[h]  
\begin{center} 
\includegraphics[height=8.0cm,width=10.0cm]{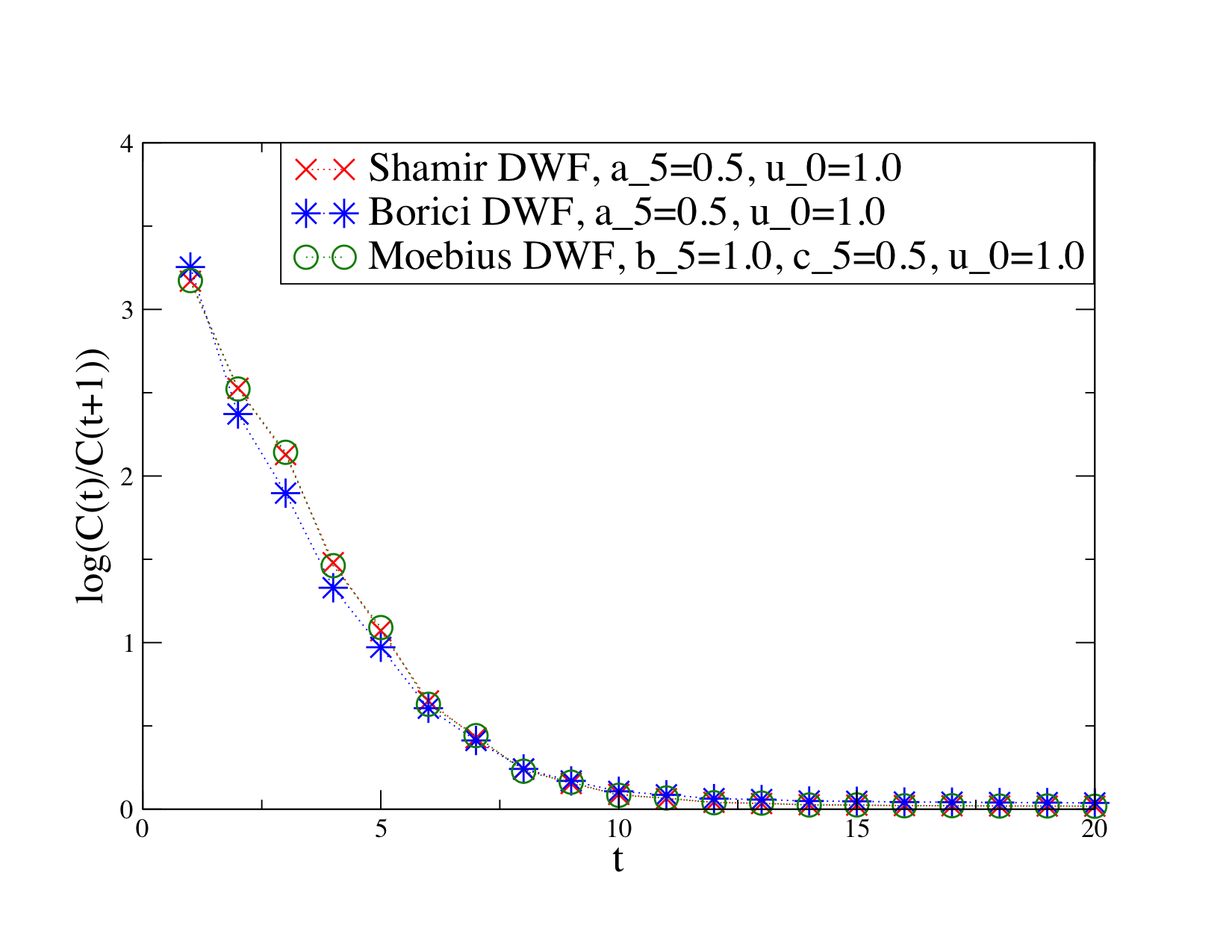}  
\caption{\small \sl $m_{eff}$ plot for $M=1.8$, $m=0.005$, $a_5=0.5$, $b_5-c_5=0.5$, $u_0=1.0$, $L_s=16.$\label{fig:2}}
\end{center}  
\end{figure}

We also verify numerically, as discussed in~\cite{shamir2}, that the oscillatory behavior reduces a lot if $a_5 \leq 0.5$, i.e. if $M a_5 <1$ and $M(b_5-a_5) <1$ in Figure~\ref{fig:2} by keeping all other parameters the same as previous plot.

An important observation to be made from from Figure~\ref{fig:1} and Figure~\ref{fig:2} is that the effective mass plot of Bori\c{c}i DWF exhibits almost zero or very little deviation from the expected exponential decay compared to Shamir and M\"{o}bius DWFs. We also point out that the effective mass plot remains the same for $a_5=1$ and $a_5=0.5$ for the Bori\c{c}i DWF in Figure~\ref{fig:2} We see that the effective mass plot is rather insensitive to change in $a_5$ for the Bori\c{c}i DWF.

\begin{figure}[h]
\begin{center} 
\includegraphics[height=8.0cm,width=10.0cm]{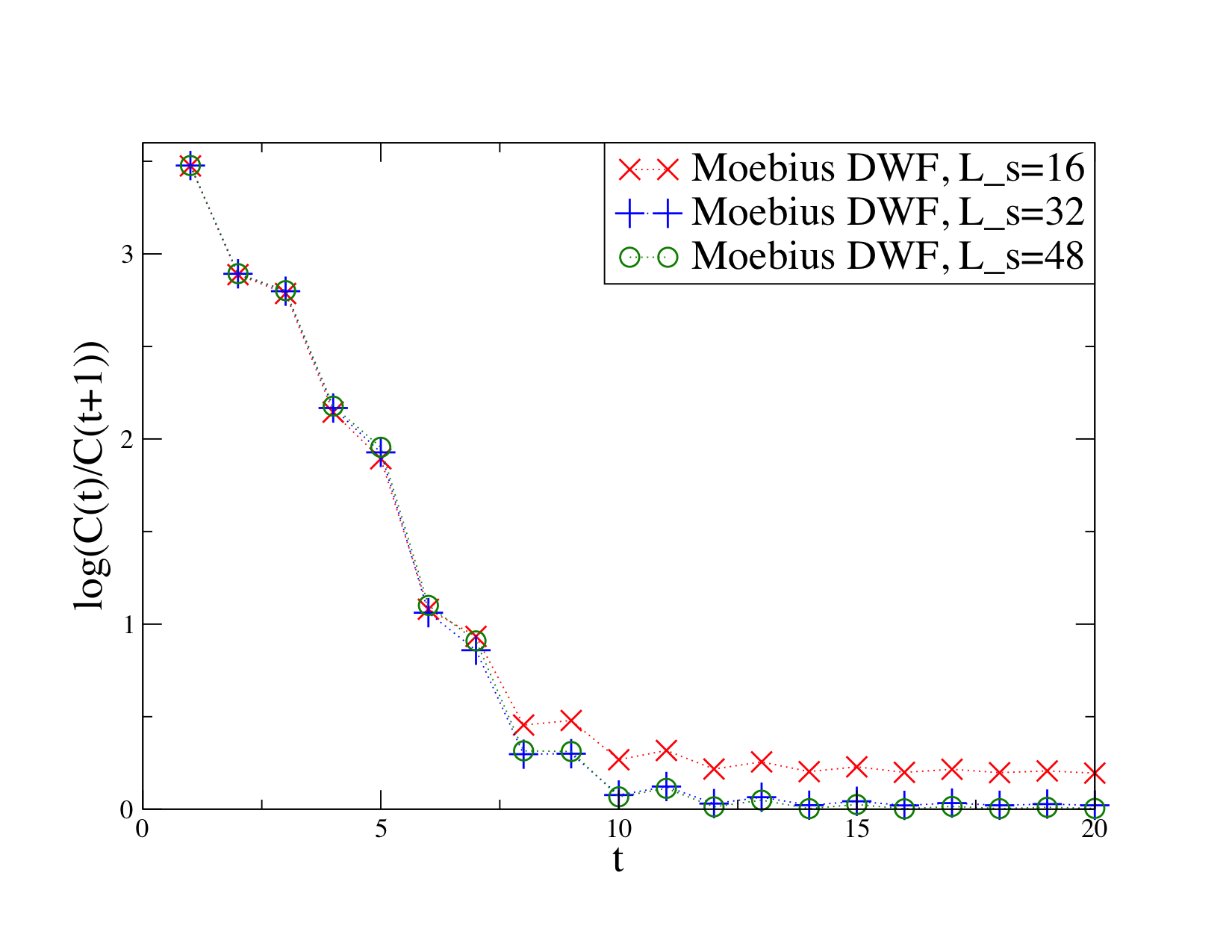}  
\caption{\small \sl $m_{eff}$ plot for M\"{o}bius DWF. $M=1.8$, $m=0.005$, $b_5-c_5=1.0$, $u_0=1.0$, $L_s=16, 32, 48.$\label{fig:3}}  
\end{center}  
\end{figure}

However, we notice from Figure~\ref{fig:1} that the plateau of the effective mass plot of the M\"{o}bius DWF does not coincide with those of Shamir and Bori\c{c}i DWF. We see, if we increase  $L_s$ from $L_s=16$ to $L_s=32$, all the plateaus coincide in the effective mass plots as shown in Figure~\ref{fig:3} for $L_s=32$.

Similar effective mass plots for different values of $L_s$ for Shamir and Bori\c{c}i DWFs can be found in Appendix~\ref{apxB}. It is interesting to notice from  Figure~\ref{fig:3} (also from Figures~\ref{fig:7},~\ref{fig:8}) that the feature of oscillation in the effective mass plots remains same if one increases $L_s$ gradually from $L_s=16$ to $L_s=48$. This is consistent with the results in Table~\ref{poles_shamir} that the unphysical pole disappears from the theory only when $L_s\to\infty$. 

\begin{figure}[h]
\begin{center} 
\includegraphics[height=8.0cm,width=10.0cm]{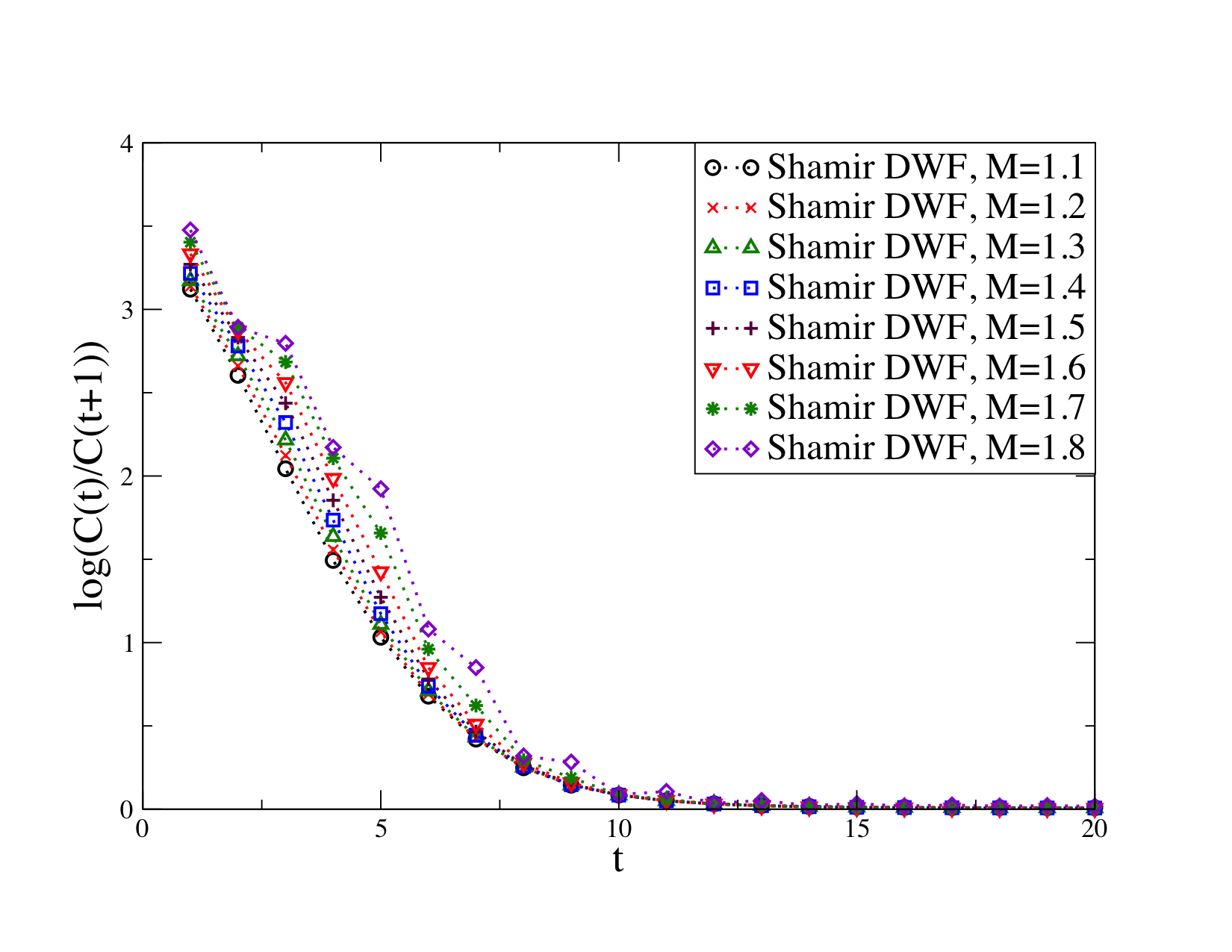}  
\caption{\small \sl $m_{eff}$ plot for Shamir DWF. $M=1.1,1.2, .....,1.8$, $m=0.005$, $a_5=1.0$, $u_0=1.0$, $L_s=16.$\label{fig:4}}  
\end{center}  
\end{figure}

\begin{figure}[h]
\begin{center} 
\includegraphics[height=8.0cm,width=10.0cm]{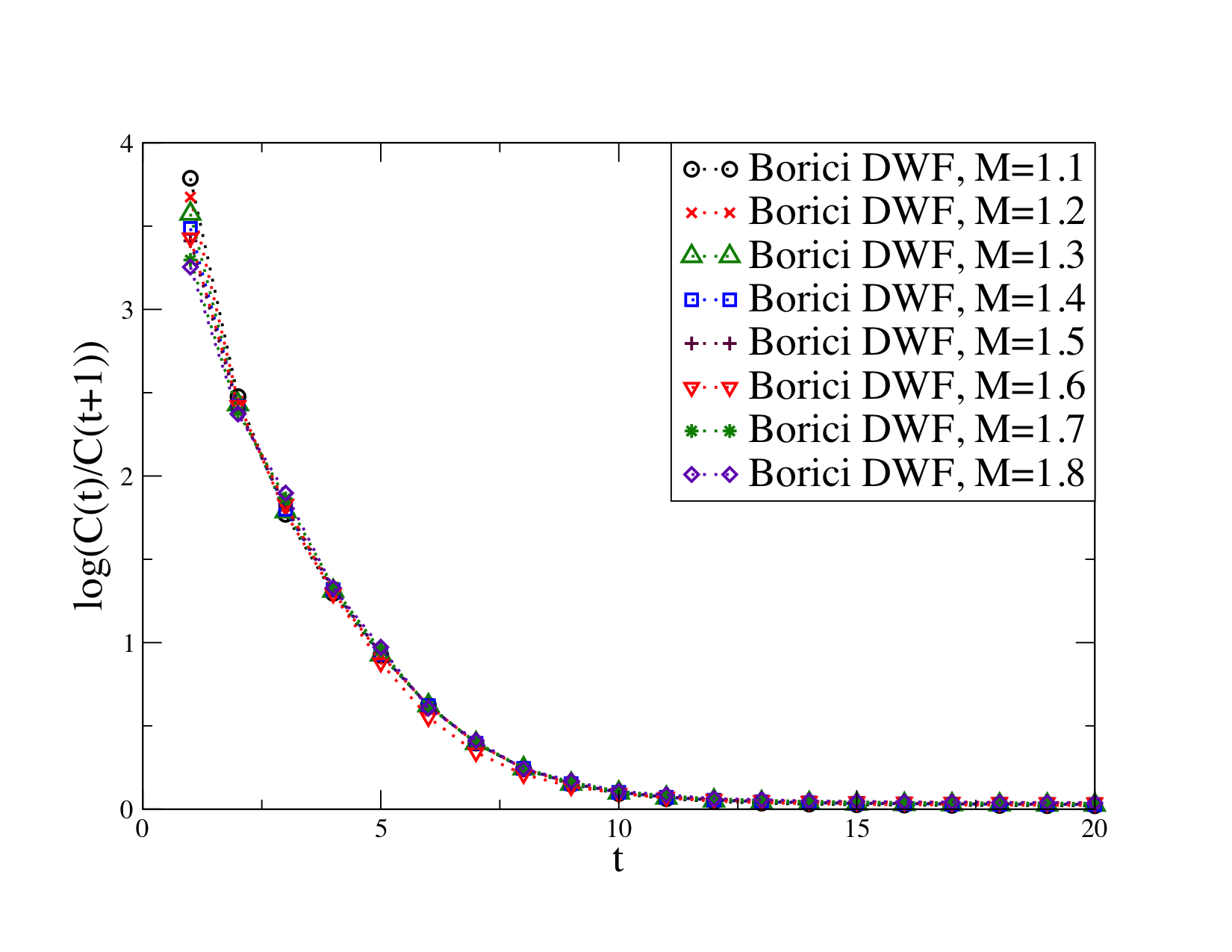}  
\caption{\small \sl $m_{eff}$ plot for Bori\c{c}i DWF. $M=1.1, 1.2, ....., 1.8$, $m=0.005$, $a_5=1.0$, $u_0=1.0$, $L_s=16.$\label{fig:5}}  
\end{center}  
\end{figure}

We now compare the effective mass plots for different values of $M=1.1, 1.2, ...., 1.8$. From Figure~\ref{fig:4} we see that at earlier time slices the value of $\log(C(t)/C(t+1))$ increases as $M$ increases for Shamir DWF.  From Figure~\ref{fig:5}, we notice at $t=1$, the value of $\log(C(t)/C(t+1))$ decreases as $M$ increases for Bori\c{c}i DWF. A similar graph for M\"{o}bius DWF has been presented in  Appendix~\ref{apxB} (Figure~\ref{fig:10}). We note that the variation of the $\log(C(t)/C(t+1))$ curves as a function of $M$ for Bori\c{c}i DWF is much smaller than both the Shamir (Figure~\ref{fig:4}) as well as the M\"{o}bius DWF (Figure~\ref{fig:10}). This indicates that the extraction of excited states might be very sensitive to the values of $M$ when Shamir and M\"{o}bius DWFs are used for numerical simulation.

\begin{figure}[h]  
\begin{center}  
\includegraphics[height=8.0cm,width=10.0cm]{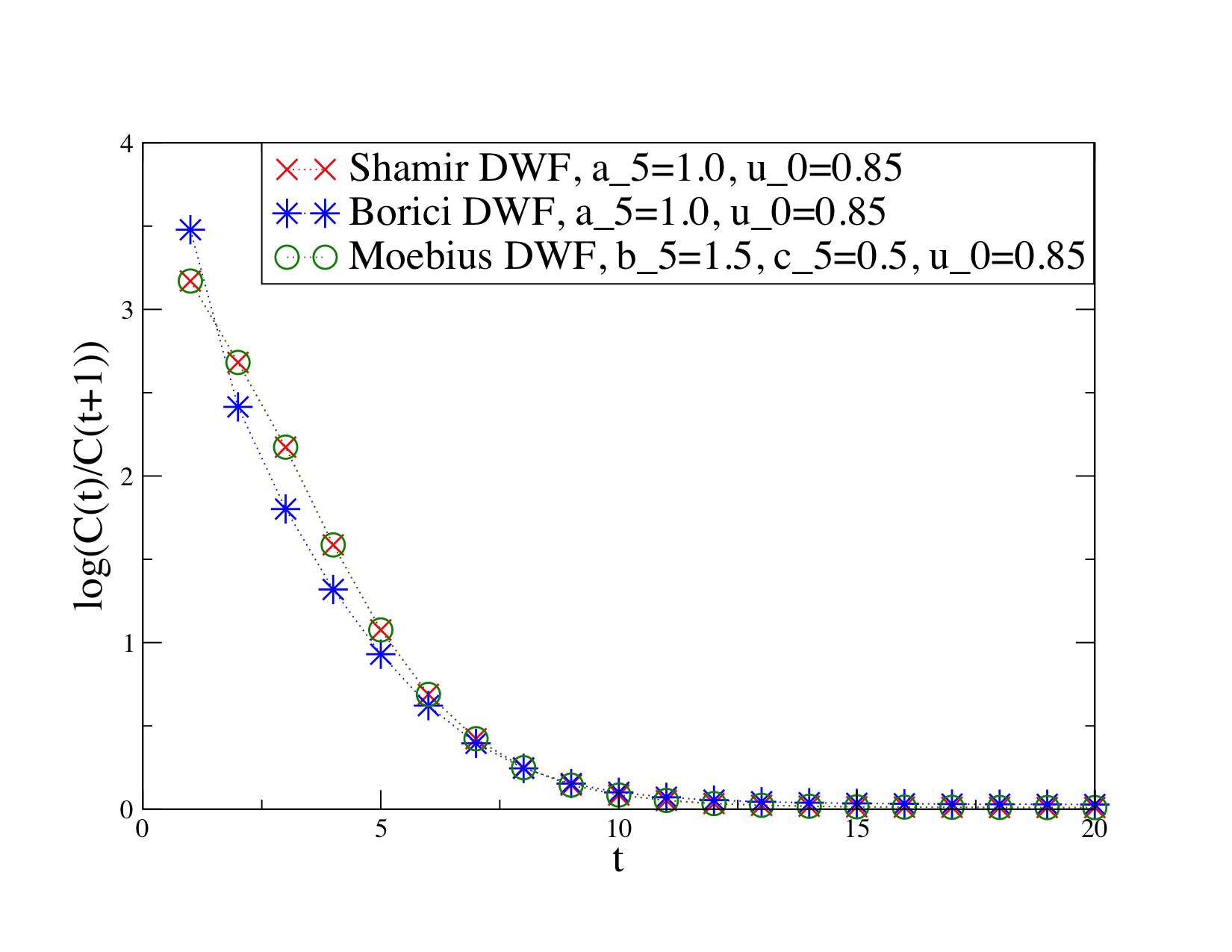}  
\caption{\small \sl $m_{eff}$ plot for $M=1.8$, $m=0.005$, $a_5=1.0$, $b_5-c_5=1.0$, $u_0=0.85$, $L_s=16.$\label{fig:6}}
\end{center}  
\end{figure}

We next consider the case where the fermion fields are coupled to a mean gauge field.  Here, the links are replaced by their vacuum expectation value, $u_0$. We choose average plaquette value $P=0.58813$~\cite{Boyle}. The mean field value of domain wall height is 
\bea
M^{MF}=M-4(1-P^{\frac{1}{4}}).
\eea
Then for $M=1.8$ the value of $M^{MF}\approx 1.302$.  $P=0.58813$ corresponds to a mean field $u_0\approx 0.85$ according to~\cite{Aoki5}. According to the discussion in~\cite{Ying}, we observe the oscillations in the effective mass plot reduce with the mean field present. This result is presented in Figure~\ref{fig:6}.  

We note that in Figures~\ref{fig:1},~\ref{fig:2},~\ref{fig:5} and~\ref{fig:6}, the oscillation in the effective mass plot for the Bori\c{c}i DWF is nearly absent.  The numerical simulations show clear evidence that Bori\c{c}i's DWF exhibits minimal deviation from exponential decay at early time slices.  However the tree level expressions for the unphysical pole are identical for every DWF action considered. To explain this fact, we consider the eigenmode expansion of the DWF propagator near the static mode (where $p_i=0$, and $p_4\neq 0$).  The poles of $(\Omega^0_{s,s'})^{-1}$ correspond to the solution $p^0_4$ which satisfies the condition,

  %Although the tree-level expression for the unphysical pole is the same for every DWF action considered, the numerical simulations show clear evidence that Bori\c{c}i's DWF exhibits minimal deviation from exponential decay at early time slices.  

\bea \label{un}
\text{Det} \, \bigg(\Omega^0_{s,s'}(p^0_4)\bigg) =0.
\eea

%Therefore it must be an eigenmode of $\Omega^0_{s,s'}(p^0_4)$ with zero eigenvalue as the zero mode corresponds to $p^{0}_4$.
For values of $p_4$ near the solution $p_4^0$, the propagator can be written as an eigendecomposition,

\bea
(\Omega^0_{s,s'})^{-1} (p_4) = \frac{1}{\lambda_0(p_4)} V^0_s(p_4) V^{0\dagger}_{s'}(p_4) + \sum_{i=1,2,..} \frac{1}{\lambda_i(p_4)}V^i_s(p_4)V^{i\dagger}_{s'} (p_4)
\eea

where $\Omega_{s,s'}^0 (p_4)V^i(p_4) = \lambda_i(p_4)V^i(p_4)$ and $\lambda_0(p_4)$ is the eigenvalue of $\Omega_{s,s'}^0 (p_4)$ which is the eigenvalue nearest zero and $V^i(p_4)$ is the corresponding eigenvector. When $p_4$ is very close to $p^0_4$, $\lambda_0(p_4)$ approaches zero and the eigenvector $V^0(p_4)$ approaches $V^0(p^0_4)$. In this case,
\bea
(\Omega^0_{s,s'})^{-1} (p_4)  \sim \frac{1}{\lambda_0(p_4^0)} V^0_s(p_4^0) V^{0\dagger}_{s'}(p_4^0)
\eea
We have shown analytically that the value of $p_4^0$ at the unphysical pole is the same for the Shamir, Bori\c{c}i and M\"{o}bius DWFs, namely $p_4^0=\ln(1-M)$. However we have observed numerically that the Bori\c{c}i DWF has the least amount of oscillation during early time slices.  

%Unlike the physical mode, the unphysical zero eigenmode corresponds to an additional propagating mode along the Euclidean time direction and is responsible for producing the oscillating mode in the temporal direction of hadron correlators.

\begin{figure}[h]  
\begin{center}  
\includegraphics[height=8.0cm,width=10.0cm]{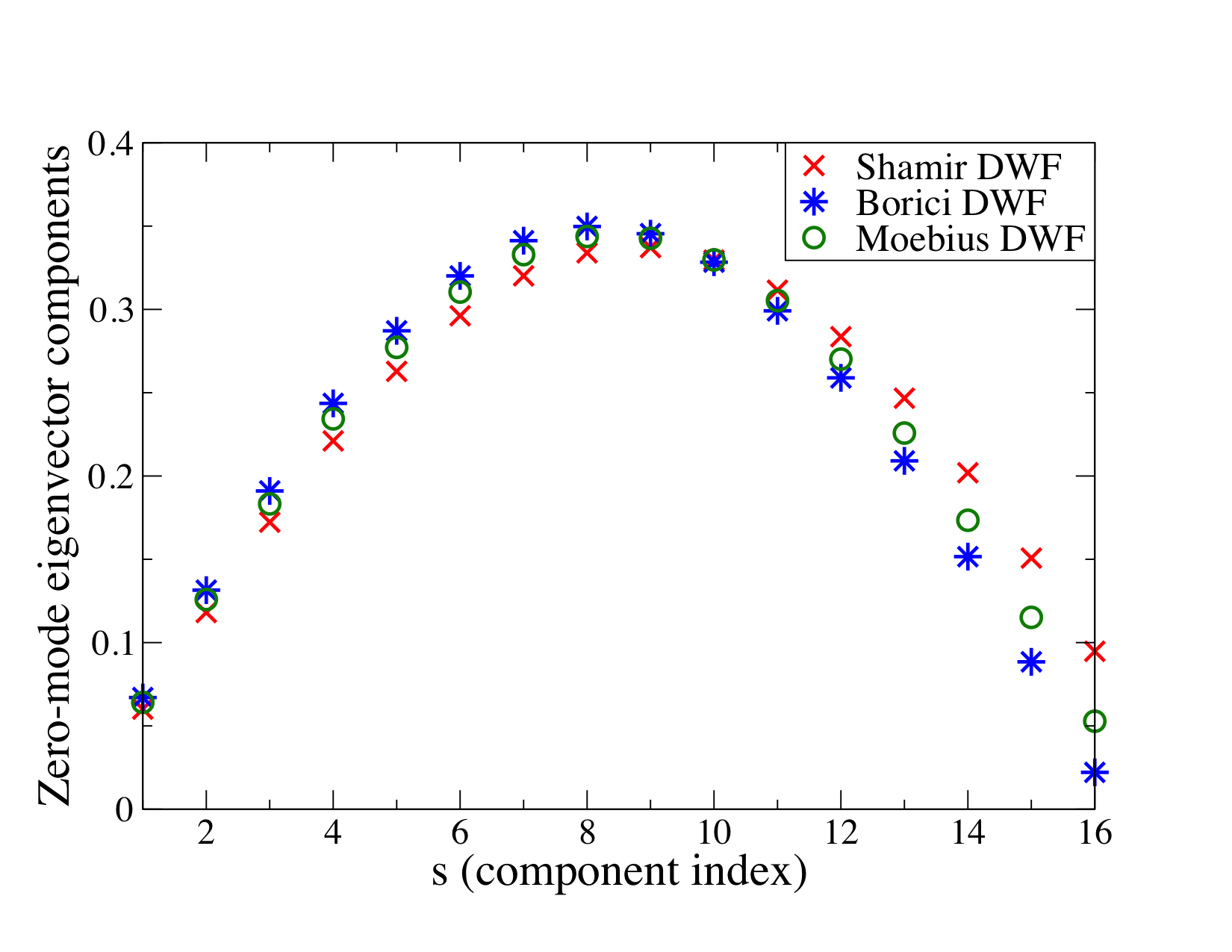}  
\caption{\small \sl Unphysical zero-mode eigenvector components contribution at 4D boundary. $M=1.3$, $m=0.001$, $a_5=1.0$, $b_5=1.5$, $c_5=0.5$, $L_s=16$. The boundary at $L_s=16$ corresponds to the 4D boundary\label{fig:eig}}
\end{center}  
\end{figure}

As discussed in~\cite{Ying}, if the contribution of the 5D unphysical eigenmode vanishes at the 4D boundary, oscillating behavior impacting 4D physics will vanish also. We illustrate this by plotting components of the zero-mode eigenvector for a value of $p_4$ corresponding to the unphysical pole (see eq.~(\ref{un})) in Figure~\ref{fig:eig} We have taken $M=1.3$, $m=0.001$, $a_5=1.0$, $b_5=1.5$, $c_5=0.5$ as parameters for the DWF transfer matrix. We observe that the $s=1$ and $s=16$ components of this eigenvector of the Bori\c{c}i DWF nearly vanish compared to the Shamir and M\"{o}bius DWFs, and therefore have minimal coupling to 4D physics. Eqs.~(\ref{q1})-(\ref{project-4d}) illustrate that the 4D propagator is the chiral projection of a product of $q(x)$ and the 5D wave function $\psi$ (s=1 or s=N). Therefore if the $s=N$ component of $\psi$ due to the unphysical mode nearly vanishes at the 4D boundary, then the unphysical modes have minimal coupling to the 4D physics. This analysis is in good agreement with our earlier $m_{eff}$ plots where we found the Bori\c{c}i DWF exhibits negligible oscillation in the hadron correlators as compared to the Shamir and M\"{o}bius DWF formalisms.

%
%%%%%%%%%%%%%%%%%%%%%%%%%%%%%%%
%  CONCLUSIONS
%%%%%%%%%%%%%%%%%%%%%%%%%%%%%%%
\section{Conclusions}\label{conclusion}
In this work, we have provided a detailed analysis of the pole structure of the free propagator for the Shamir, Bori\c{c}i, and M\"{o}bius DWF actions.  By investigating the poles we have provided a qualitative picture of the origin of oscillatory behavior in the effective mass plots when using DWFs and have quoted restrictions on the values of the product $M a_5<1$ (Shamir, Bori\c{c}i), $M(b_5-c_5)<1$ (M\"{o}bius) needed to ensure such oscillatory behavior is very small or absent for $M>1$. We have also pointed out that the Bori\c{c}i DWF correlators are insensitive to the change of values of $a_5$.  We have shown for free case that at earlier time $\log(C(t)/C(t+1))$ curves for Shamir and M\"{o}bius DWFs depend significantly on the values of $M$. Therefore one needs to find out a value of $M$ such that extraction of hadron excited states produces the correct result. 

In all cases, we observe that the Bori\c{c}i DWF has far less oscillatory effects at early time slices as compared to the other DWF actions we consider. We have argued why this is so by considering the contribution of unphysical eigenmodes on the 4D boundary of Bori\c{c}i DWF.  We have shown that these eigenmodes are suppressed on the 4D boundary by a larger amount compared to Shamir and M\"{o}bius actions, which corresponds to smaller oscillatory effects contaminating the 4D physics of interest.
  
We have commented that the effects of the unphysical pole cannot be removed from the transfer matrix when realistic simulations are performed at a finite $L_s$.  Because of this, the unphysical mode can have a significant impact on any fermion loop calculation when DWFs are employed. We close this section with the remark that even though this analysis represents a non-interacting analysis of the DWF pole structure,  we expect similar qualitative features to be present in fully interacting calculations.   

%%%%%%%%%%%%%%%%%%

\appendix
\section{Appendix A}\label{apxA}
The 4D Bori\c{c}i propagator in the momentum space:
\bea
S^{4D}(p)&&= i\slashed{\overline{p}}\Bigg(-\frac{m'^2_r((c^2+\overline{p}^2)(1-b'e^{-\alpha'}) -m'^2(1-be^{\alpha'})) }{B'F'_N}+\frac{2m'^2 m'_r}{F'_N}+\nn \\
&&\frac{m'^2_r(c^2+\overline{p}^2-m'^2)(1-be^{\alpha'})}{B'F'_N}-\frac{1}{B'}-\frac{(c^2+\overline{p}^2-m'^2)(1-b'e^{-\alpha'})}{B'F'_N}\nn \\
&&+\frac{m'^2_r ((c^2+\overline{p}^2) (1-b'e^{-\alpha'})-m'^2(1-b'e^{\alpha'})  )}{B'F'_N}\Bigg) -b'(p)\Bigg( \frac{m'_re^{\alpha}}{B'}-\frac{m'(m'^2_re^{\alpha}+e^{-\alpha})}{F'_N}\nn \\
&& -\frac{m'_re^{-\alpha}(c^2+\overline{p}^2-m'^2)(1-b'e^{\alpha})}{B'F'_N}+\frac{m'^3_re^{\alpha}-m'_re^{-\alpha}((c^2+\overline{p}^2)(1-b'e^{-\alpha})-m'^2(1-b'e^{\alpha'}))}{B'F'_N}\Bigg) \nn \\
&&+m'\Bigg(\frac{m'^2_r((c^2+\overline{p}^2)(1-b'e^{-\alpha'}) -m'^2(1-be^{\alpha'})) }{B'F'_N}-\frac{m'^2_r(c^2+\overline{p}^2-m'^2)(1-be^{\alpha'})}{B'F'_N}+\frac{1}{B'}\nn \\
&&-\frac{2m'^2 m'_r}{F'_N}
-\frac{(c^2+\overline{p}^2-m'^2)(1-b'e^{-\alpha'})}{B'F'_N}-\frac{m'^2_r ((c^2+\overline{p}^2) (1-b'e^{-\alpha'})-m'^2(1-b'e^{\alpha'})  )}{B'F'_N}\Bigg)  \nn \\
\eea
where in the limit $a,a_5\to 1$ we have defined,
\bea
b' &&=\left[-\overline{p}^{2}-b(p)c(p)\right] \nn \\
m'&&=-m(D_w(p)-1) \nn \\
F_N&&=(c^2(p)+\overline{p}^2)(1-b'e^{\alpha'})-m'^2(1-b'e^{-\alpha'})\nn \\
&&+m'_r[4\sinh\alpha'(c^2(p)+\overline{p}^2-m'^2)]\nn \\
&&-m'^2_r[(c^2(p)+\overline{p}^2)(1-b'e^{-\alpha'})-m'^2(1-b'e^{\alpha'})] \nn \\
m'^2&&=m^2(c^2(p)+\overline{p}^2)\nn \\
m'_r&&=e^{-\alpha' L_s}
\eea
\section{Appendix B}\label{apxB}
\begin{figure}[h]
\begin{center} 
\includegraphics[height=8.0cm,width=10.0cm]{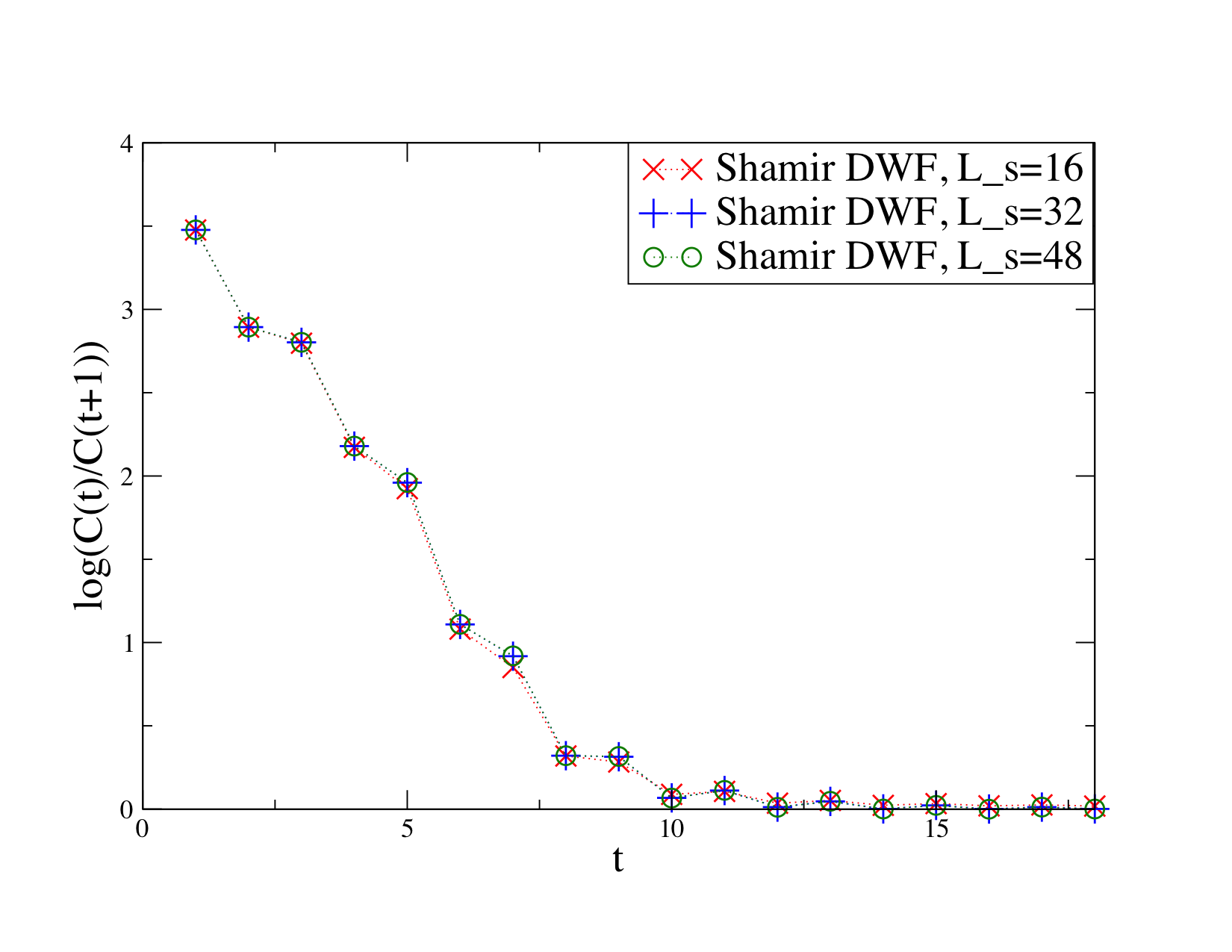}  
\caption{\small \sl $m_{eff}$ plot for Shamir DWF. $M=1.8$, $m=0.005$, $a_5=1.0$, $u_0=1.0$, $L_s=16, 32, 48.$\label{fig:7}}  
\end{center}  
\end{figure}

\begin{figure}[h]
\begin{center} 
\includegraphics[height=8.0cm,width=10.0cm]{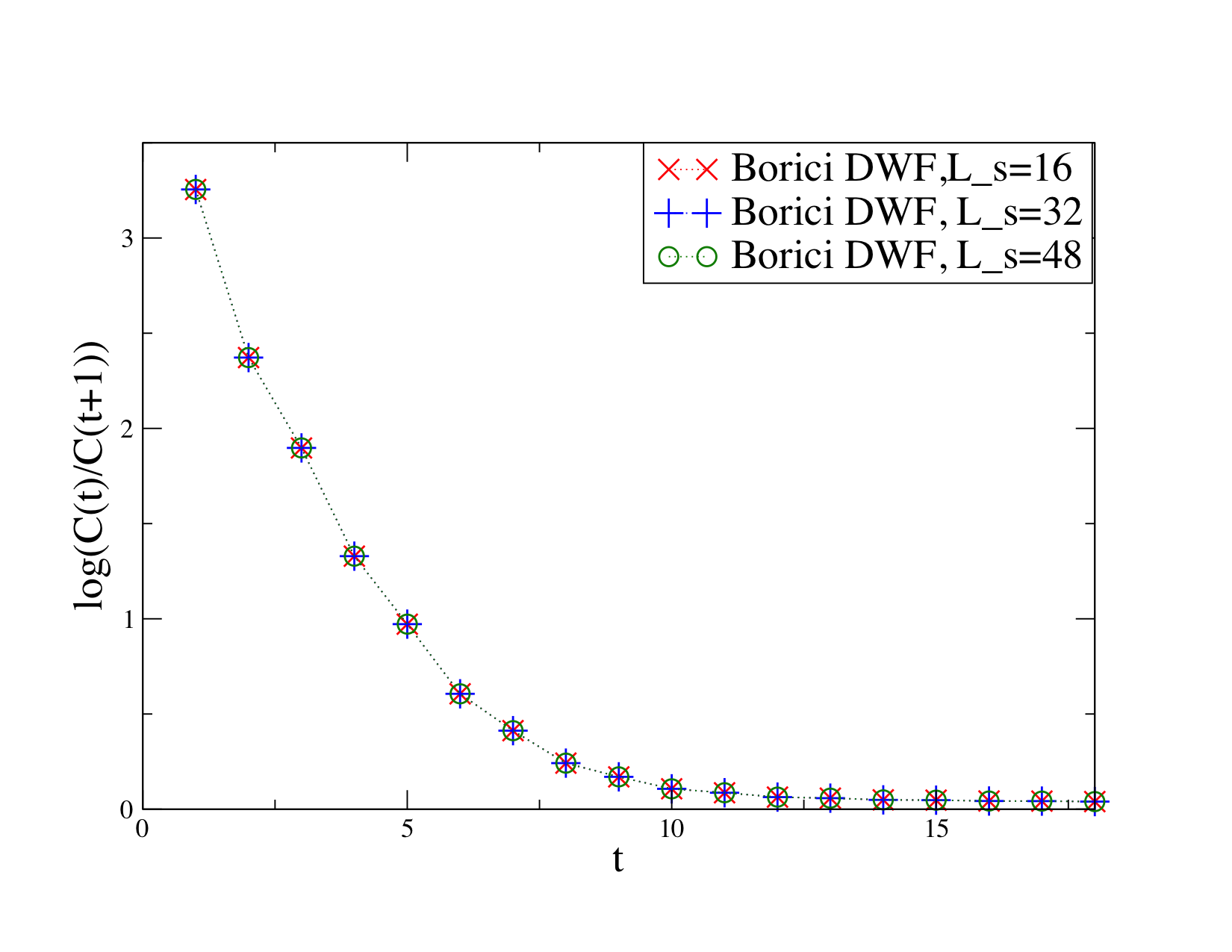}  
\caption{\small \sl $m_{eff}$ plot for Bori\c{c}i DWF. $M=1.8$, $m=0.005$, $a_5=1.0$, $u_0=1.0$, $L_s=16, 32, 48.$\label{fig:8}}  
\end{center}  
\end{figure}

\begin{figure}[h]
\begin{center} 
\includegraphics[height=8.0cm,width=10.0cm]{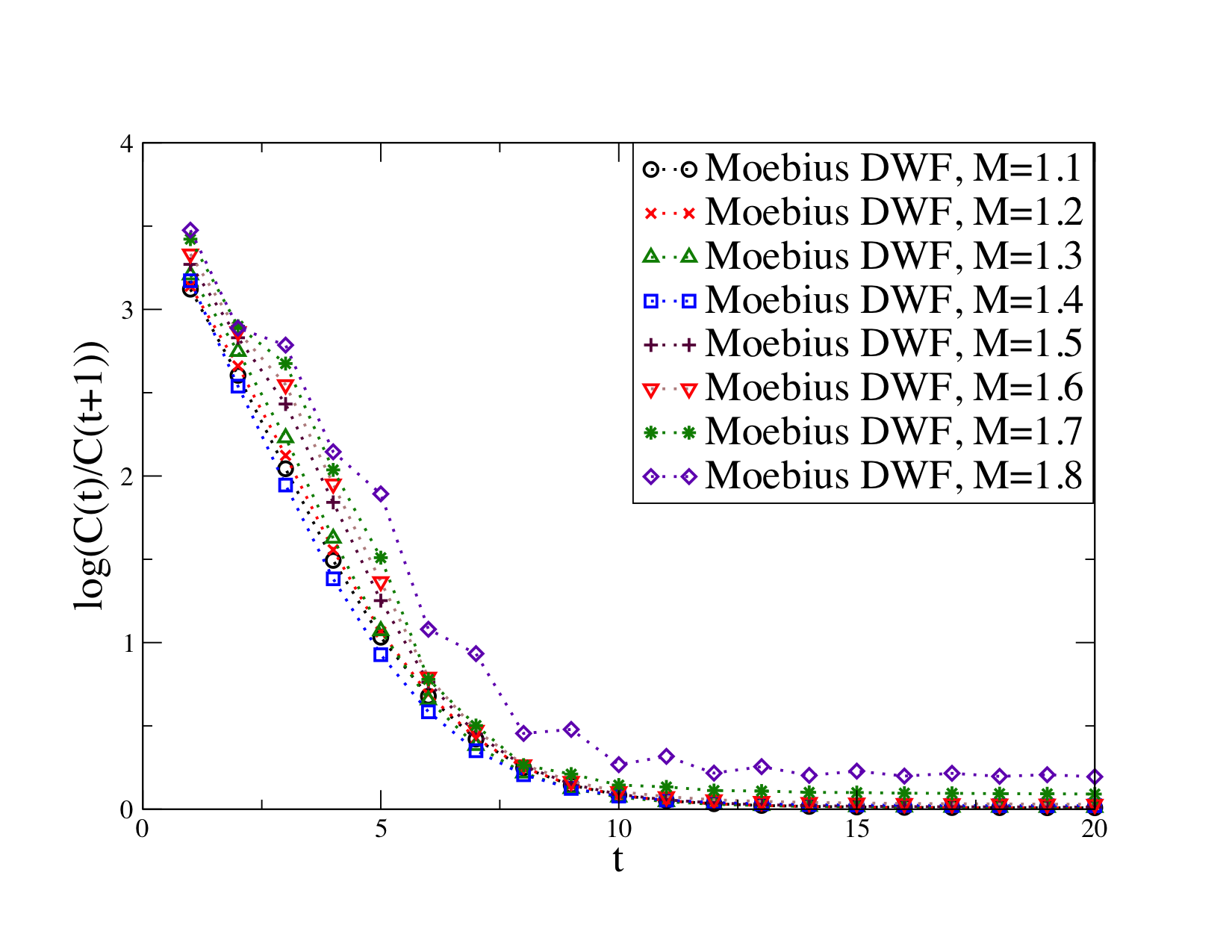}  
\caption{\small \sl $m_{eff}$ plot for M\"{o}bius DWF. $M=1.1, 1.2, ...., 1.8$, $m=0.005$, $b_5-c_5=1.0$, $u_0=1.0$, $L_s=16.$\label{fig:10}}  
\end{center}  
\end{figure}

\acknowledgments

The authors thank Keh-Fei Liu who provided insight and expertise that greatly assisted the research. R.S.S. thanks Sergey Syritsyn and Ying Chen for their valuable suggestions throughout the course of this work. 
%\paragraph{Note added.} This is also a good position for notes added
%after the paper has been written.


\end{document}